\documentclass[twocolumn,aps,prl,preprintnumbers]{revtex4}
\usepackage{bbm}
\usepackage[latin9]{inputenc}
\usepackage{amsmath}
\usepackage{amssymb}
\usepackage{graphicx}
\usepackage{mathrsfs}
\usepackage{amsfonts}
\usepackage{amsthm}
\usepackage{color}
\usepackage{txfonts}
\usepackage{lipsum}
\usepackage[colorlinks=true,citecolor=blue,linkcolor=blue,urlcolor=blue,anchorcolor=blue]{hyperref}%
\hypersetup{colorlinks=true,citecolor=blue,linkcolor=blue,urlcolor=blue}
\providecommand{\U}[1]{\protect\rule{.1in}{.1in}}
\makeatletter
\@ifundefined{textcolor}{}
{
\definecolor{BLACK}{gray}{0}
\definecolor{WHITE}{gray}{1}
\definecolor{RED}{rgb}{1,0,0}
\definecolor{GREEN}{rgb}{0,1,0}
\definecolor{BLUE}{rgb}{0,0,1}
\definecolor{CYAN}{cmyk}{1,0,0,0}
\definecolor{MAGENTA}{cmyk}{0,1,0,0}
\definecolor{YELLOW}{cmyk}{0,0,1,0}
}
\makeatother
\begin{document}
\title{Exceptional magnetic sensitivity of $\mathcal {P}\mathcal {T}-$symmetric cavity magnon polaritons}
\author{Yunshan Cao}
\email[]{yunshan.cao@uestc.edu.cn}
\author{Peng Yan}
\email[]{yan@uestc.edu.cn}
\affiliation{School of Electronic Science and Engineering and State Key Laboratory of Electronic Thin Films and Integrated Devices, University of
Electronic Science and Technology of China, Chengdu 610054, China}

\begin{abstract}
Achieving magnetometers with ultrahigh sensitivity at room temperature is an outstanding problem in physical sciences and engineering. Recently developed non-Hermitian cavity spintronics offers new possibilities. In this work, we predict an exceptional magnetic sensitivity of cavity magnon polaritons with the peculiar parity-time ($\mathcal {P}\mathcal {T}$) symmetry. Based on the input-output formalism, we demonstrate a $``Z"-$shape spectrum including two side-band modes and a dark-state branch with an ultra-narrow linewidth in the exact $\mathcal {P}\mathcal {T}$ phase. The spectrum evolves to a step function when the polariton touches the third-order exceptional point, accompanied by an ultrahigh sensitivity with respect to the detuning. The estimated magnetic sensitivity can approach $10^{-15}\ \mathrm{T}\ \mathrm{Hz}^{-1/2}$ in the strong coupling region, which is two orders of magnitude higher than that of the state-of-the-art magnetoelectric sensor. We derive the condition for the noise-less sensing performance. Purcell-like effect is observed when the $\mathcal {P}\mathcal {T}$ symmetry is broken. Possible experimental scheme to realize our proposal is also discussed.
\end{abstract}
\maketitle

\begin{center}
\textbf{I. INTRODUCTION}
\end{center}

Strong light-matter interaction lies in the heart of cavity (or circuit) quantum electrodynamics (CQED) and quantum information science. It allows the Rabi splitting and polaritonic eigenmodes. The subject has been extensively studied in the hybridized cavity and two-level system, including atoms \cite{Walther2006}, molecules \cite{Shalabney2015,Chikkaraddy2016}, superconducting qubits \cite{Wallraff2004}, and quantum dots \cite{Yoshie2004,Gross2018}. In recent years, cavity spintronics (or spin cavitronics)---the emerging interdiscipline of CQED and spintronics---has been rapidly developing \cite{Huebl2013,Bhoi2014,Tabuchi2014,Zhang2014,Goryachev2014,Zhang2015,Bai2015,Flaig2016}. A central issue in the community is to observe the cavity magnon polariton (CMP)---the hybrid quasiparticle of the microwave photon coherently coupled with magnon (or spin wave), the collective excitation in ordered magnets that can efficiently interact with external magnetic fields. The entangled spin orientation and photon number state enables an efficient quantum information transfer between photon and magnon via Rabi oscillation, which is promising for quantum computing \cite{Washington2018}. Thanks to the extremely low damping and high spin density in ferrimagnetic insulators like yttrium iron garnet (YIG) \cite{Wu2013}, the CMPs have been observed at both cryogenic temperature and room temperature from the level repulsion spectra \cite{Huebl2013,Bhoi2014,Tabuchi2014,Zhang2014,Goryachev2014,Zhang2015,Bai2015,Flaig2016,Cao2015,Rameshti2015,Tabuchi2015,Quirion2017}. Very recently, an exotic level attraction of non-Hermitian magnon-photon coupling was reported \cite{Harder2018,Grigoryan2018}, which opens a new avenue for exploring spin cavitronics.

On the other hand, highly sensitive magnetometers are indispensable tools which assisted humankind through a wide range of practical applications in geology, navigation, archaeology, magnetic storage, and medicine \cite{Lenz2006,Edelstein2007,Grosz2017}. The technologies used for magnetic field sensing encompass many aspects of physics, such as search coil, fluxgate, Hall effect,
magnetoelectric coupling, spin mechanics, and magnetoresistance \cite{Bichurin2002,Annapureddy2017,Kimball2016,Band2018,Reig2013}, to name a few. The state-of-the-art magnetic sensor can reach an ultrahigh sensitivity of subfetotesla, like a superconducting quantum interference device \cite{Gallop2003,Kleiner2004} and atomic magnetometer \cite{Dang2010,Kominis2003}, however with limitations such as extreme temperature \cite{Gallop2003,Kleiner2004} or low working frequency \cite{Kominis2003,Dang2010}. \begin{figure}
  \centering
  \includegraphics[width=0.8\linewidth]{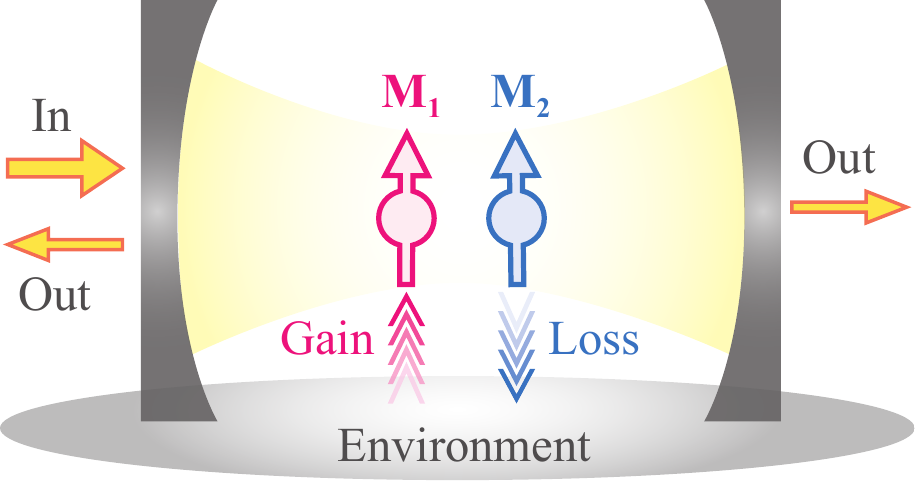}\\
  \caption{Schematic illustration of photon scattering by two magnons with balanced gain and loss in a microwave cavity.}\label{system}
\end{figure}
Pursuing solid-state room-temperature magnetometers with exceptional sensitivity represents a critical and challenging problem. Despite working in very different contexts, the magnetometers mentioned above share the same basic principle: at a conventional Hermitian degeneracy (also called diabolic point), the induced shift of any physical quantity (e.g. the magnetoresistance) by external perturbation $\epsilon$ (e.g. the magnetic field) is linear with the perturbation itself (with $|\epsilon|\ll 1$). This rule, however, is broken for a non-Hermitian degeneracy called exceptional point (EP) at which not only the eigenvalues, but also the eigenstates are simultaneously coalesced. The most exciting non-Hermitian system are those respecting parity-time ($\mathcal {P}\mathcal {T}$) symmetry which could exhibit entirely real spectra below the EP \cite{Bender1998,Bender2002}. The order of EP is determined by the number of degenerate eigenstates.
Non-Hermitian perturbation theory shows that the eigenfrequency shift follows a $|\epsilon|^{1/N}-$dependence at the $N$-th order EP, with $N$ an integer. The non-Hermitian degeneracy thus can significantly enhance the sensitivity \cite{Wiersig2014}.
Currently, $\mathcal {P}\mathcal {T}$ symmetry has been investigated in a broad field of quantum mechanics \cite{Bender1998,Bender2002,Bender2007}, optics \cite{Feng2013,Feng2014,Peng2014,Konotop2016,Wen2018}, acoustics \cite{Zhu2014,Jing2014}, electronics \cite{Schindler2011,Schindler2012,Chen2018}, and very recently in spintronics \cite{Harder2017,Zhang2017} and magnonics \cite{Lee2015,Galda2016,Galda2018,Yang2018}. However, the property of $\mathcal {P}\mathcal {T}$-symmetry and EP sensing are yet to be addressed in spin cavitronics.

In this work, we theoretically study the non-Hermitian coupling between a cavity photon and two magnons with the $\mathcal {P}\mathcal {T}$ symmetry (see Fig. \ref{system}). We predict an exceptional magnetic sensitivity around the third-order EP of CMP. From the input-output formalism, we analytically derive the transmission coefficient $S_{21}$ and identify a novel $``Z"-$shape spectrum in the exact $\mathcal {P}\mathcal {T}$ phase without avoided level crossing. The spectrum evolves to a step function when the polariton touches the third-order EP, exhibiting an ultrahigh sensitivity with respect to the detuning. The estimated magnetic sensitivity approaches two orders of magnitude higher than that of the state-of-the-art magnetoelectric sensor. Practical realization of our proposal is discussed.

The paper is organized as follows. Theoretical model is presented in Sec. II. Section III gives the main results, including the phase diagram, the transmission spectrum, the magnetic sensitivity, and the nonlinear effect. We implement a classical wave scattering calculation in Sec. IV. Discussion and conclusion are drawn in Sec. V and Sec. VI, respectively. Formula derivation is elaborated in the Appendix.

\begin{center}
\textbf{II. MODEL}
\end{center}

We consider a setup consisting of two magnetic bodies with balanced gain and loss inside a microwave cavity (shown in Fig. \ref{system}). The magnetization is connected to the local spin operator via $\mathbf{M}=-\gamma \mathbf{S}$ with $\gamma$ the gyromagnetic ratio. Considering small amplitude excitations and using the Holstein-Primakoff transformation, we can write the spin operator as $S_z\sim -{\hat{s}^\dagger}\hat{s} $ and $S_{x,y}\sim (\hat{s}^\dagger\pm \hat{s})$. We consider the following non-Hermitian bosonic Hamiltonian
\begin{eqnarray}\label{Eq-Ham}
  \mathcal{H}&=&\hbar\omega_\mathrm{c}{\hat{a}^\dagger}\hat{a}
  +\hbar(\omega_\mathrm{s}+i\beta){\hat{s}^\dagger_1}\hat{s}_1+\hbar(\omega_\mathrm{s}-i\beta){\hat{s}^\dagger_2}\hat{s}_2\\ \nonumber
  &&+\hbar g \left[{\hat{a}^\dagger}(\hat{s}_1+\hat{s}_2)+h.c.\right],
\end{eqnarray}
where $\hat{a}^\dagger(\hat{a})$ and $\hat{s}^\dagger_{1,2}(\hat{s}_{1,2})$ are the photon and magnon creation
(annihilation) operators, respectively, $\omega_\mathrm{c}$ is the cavity resonant frequency, $\omega_\mathrm{s}$ denotes the Zeeman splitting, $\beta>0$ describes the energy dissipation/amplification rate with environments, and $g$ represents the magnon-photon coupling strength. Under a combined operation of parity $\mathcal {P}$ ($\hat{s}_1\leftrightarrow \hat{s}_2$) and time reversal $\mathcal {T}$ ($i\rightarrow-i$, $\hat{s}_{1(2)}\rightarrow-\hat{s}_{1(2)}$, and $\hat{a}\rightarrow-\hat{a}$), it is straightforward to find that Eq. (\ref{Eq-Ham}) is invariant and thus respects the $\mathcal {P}\mathcal {T}$ symmetry. The direct exchange coupling between magnons is assumed to be absent in the present model. We note that Hamiltonian (\ref{Eq-Ham}) is a non-trivial generalization of the purely dissipative one adopted in Ref. \cite{Plenio1999}.

We consider single particle processes, so that three states $\left\{\hat{a}^\dagger|0\rangle, ~\hat{s}^\dagger_1|0\rangle, ~\hat{s}^\dagger_2|0\rangle\right\}$ constitute the complete basis, where $|0\rangle$ represents the vacuum state. The Hamiltonian can therefore be expressed in the following matrix form (set $\hbar=1$),
\begin{equation}
\mathcal{H}=\left(
      \begin{array}{ccc}
        \omega_\mathrm{s}+i\beta & 0 & g \\
        0 & \omega_\mathrm{s}-i\beta & g \\
        g & g & \omega_\mathrm{c} \\
      \end{array}
    \right).
\end{equation}
By solving $\mathcal{H}|\phi\rangle=\omega|\phi\rangle$, we obtain the following cubic equation for the eigenvalues,
\begin{equation}\label{Eq-Cubic}
\left(\Omega^2+P^2\right)\left(\Omega+\Delta\right)-2\Omega=0,
\end{equation}
with $\Omega=(\omega-\omega_\mathrm{s})/g$, $\Delta=(\omega_\mathrm{s}-\omega_\mathrm{c})/g$ the frequency detuning, and $P=\beta/g$ being the balanced gain-loss parameter.

\begin{center}
\textbf{III. RESULTS}
\end{center}

\begin{center}
\textbf{A. Phase diagram}
\end{center}

\begin{figure}
  \centering
  \includegraphics[width=\linewidth]{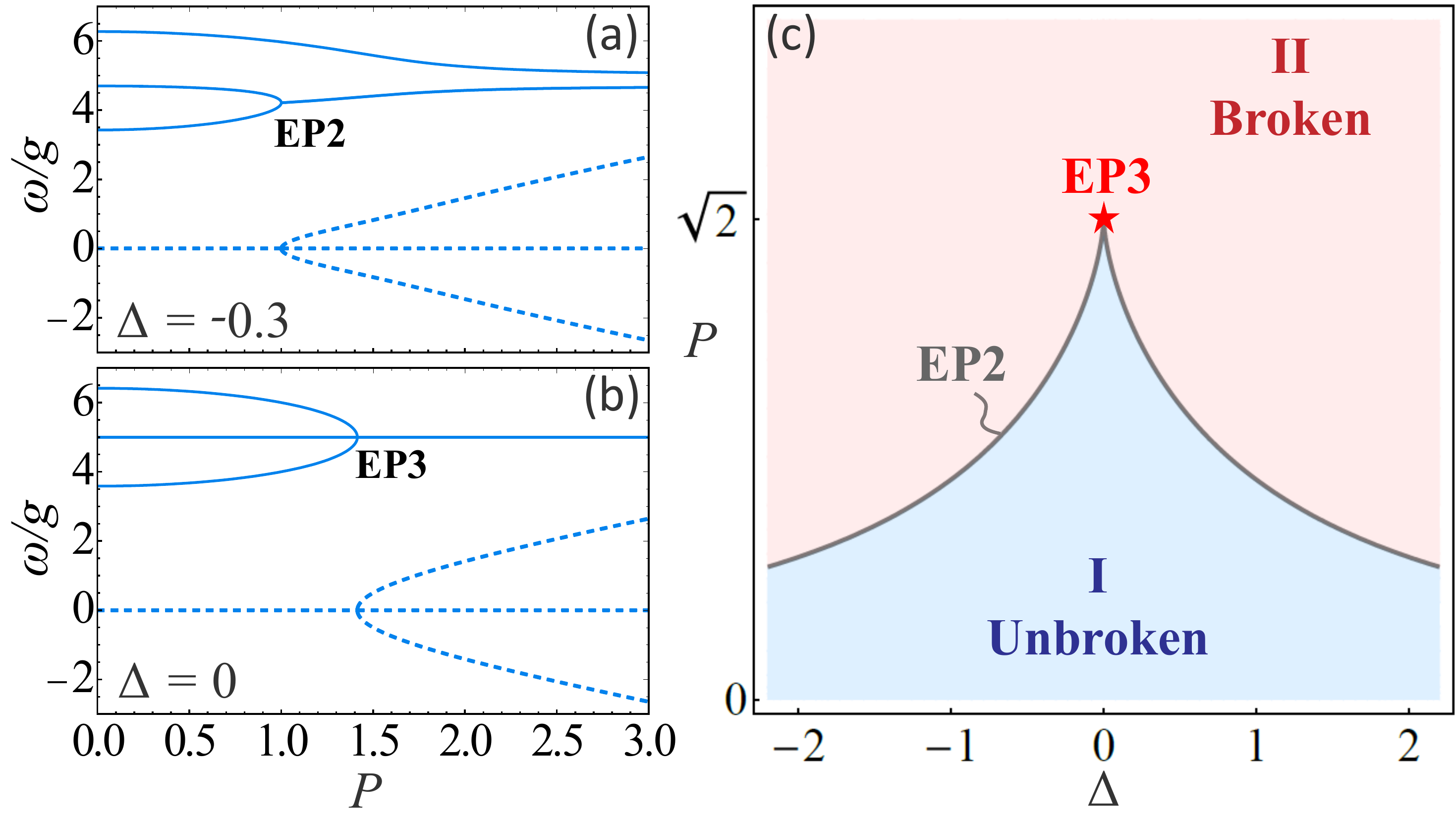}\\
  \caption{Evolution of eigenvalues as the gain-loss parameter $P$, with the solid and dashed curves respectively representing the real and imaginary part of eigenfrequencies.
  The detuning parameters are chosen to be (a) $\Delta=-0.3$ and (b) $\Delta=0$.
  The cavity frequency is set as $\omega_\mathrm{c}/g=5$.
  (c) $\mathcal {P}\mathcal {T}$-symmetric phase transition diagram.}\label{phase}
\end{figure}

Figure \ref{phase}(a) shows the roots of (\ref{Eq-Cubic}) with a detuning parameter $\Delta=-0.3$. There are three real solutions at a small $P$, which corresponds to the unbroken $\mathcal {P}\mathcal {T}$ phase. By increasing $P$, one pair of eigenvalues coalesce at $P_\mathrm{EP2}$ and then bifurcate into the complex plane when $P>P_\mathrm{EP2}$. Here EP$N$ represents the $N$th-order EP. For a zero detuning ($\Delta=0$), the closed-form solutions of the three eigenvalues are $\omega=\omega_\mathrm{c}\pm\sqrt{2g^2-\beta^2}$ for side modes and $\omega=\omega_\mathrm{c}$ for the central mode, as shown in Fig. \ref{phase}(b). The third-order exceptional point EP3 appears when $\beta=\sqrt{2}g$ (or $P=P_\mathrm{EP3}=\sqrt{2}$), with the unique coalesced eigenstate being $(\frac{i}{2},\frac{-i}{2},\frac{1}{\sqrt{2}})^{\text{T}}$. For $P<P_\mathrm{EP3}$, the side modes exhibit an abnormal Rabi splitting with the frequency separation depending not only on the coupling strength, but also on the gain-loss parameter. This is a sharp contrast to their Hermitian counterpart. Further, we note that the flat central mode (real for all $P$) actually corresponds to a dark-state polariton \cite{Fleischhauer2000,Dong2012,Emary2013} (see analysis below).

The phase diagram [plotted in Fig. \ref{phase}(c)] is determined by the sign of the discriminant
\begin{equation}
\Lambda=P^2\Delta^4+(2P^4+10P^2-1)\Delta^2+(P^2-2)^3
\end{equation}
of (\ref{Eq-Cubic}). $\Lambda<0$ gives the exact (or unbroken) $\mathcal {P}\mathcal {T}$ phase, in which all three eigenvalues are real and the eigenvectors satisfy the so-called biorthogonal relation $\langle\phi_i^*|\phi_j\rangle=\delta_{ij}$ with $i,j=1,2,3$ \cite{Brody2014}. For $\Lambda>0$, only one real eigenvalue survives and the other two become complex conjugated, which corresponds to the broken $\mathcal {P}\mathcal {T}$ phase. EP2 happens along the critical curve $\Lambda=0$ but with $\Delta\neq0$ [see the grey curve in Fig. \ref{phase}(c)]. EP3 emerges when both $P=P_\text{EP3}=\sqrt{2}$ and $\Delta=0$ are simultaneously satisfied [see the red star in Fig. \ref{phase}(c)].

\begin{center}
\textbf{B. Transmission spectrum }
\end{center}

For a conventional hybridized CMP system, the strong coupling is usually identified from the gap of the transmission spectrum at the resonance point. Next, we derive the scattering coefficient of the $\mathcal{PT}$-symmetric CMP system via a standard input-output theory. We assume that the cavity is interacting with a harmonic bosonic bath (environment). By introducing the noise and dissipation functions into the Heisenberg equations of operators, we obtain the following quantum Langevin equations \cite{Lax1966,Gardiner1985},
\begin{subequations}\label{Eq-Lang}
\begin{eqnarray}
    &&\dot{\hat{a}}=(-i\omega_\mathrm{c}-\kappa_\mathrm{c})\hat{a}-ig(\hat{s}_1+\hat{s}_2)-\sqrt{\kappa_\mathrm{c}}\hat{b}_\mathrm{in},\\
    &&\dot{\hat{s}}_1 =(-i\omega_\mathrm{s}+\beta)\hat{s}_1-ig\hat{a}, \hspace{5em}\mathrm{(gain)}\\
    &&\dot{\hat{s}}_2 =(-i\omega_\mathrm{s}-\beta)\hat{s}_2-ig\hat{a}, \hspace{5em}\mathrm{(loss)}
\end{eqnarray}
\end{subequations}
where $\kappa_\mathrm{c}$ represents the leakage rate of a photon to the environment (the internal loss of the cavity is assumed to be negligibly small), and $\hat{b}_{\mathrm{in}/\mathrm{out}}$ is the input/output field from the thermal bath, satisfying the input-output formula $\hat{b}_\mathrm{out}=\hat{b}_\mathrm{in}+2\sqrt{\kappa_\mathrm{c}} \hat{a}$ \cite{Gardiner1985,Meystre2007,Walls2008}. After some algebra, we obtain the frequency-resolved transmission coefficient (see Appendix A for details),
\begin{equation}\label{Eq-S21}
    S_{21}=\frac{\kappa_\mathrm{c}}{i(\omega-\omega_\mathrm{c})-\kappa_\mathrm{c}+\Sigma(\omega)},
\end{equation}
where the total self-energy $\Sigma(\omega)=\Sigma^\mathrm{gain}(\omega)+\Sigma^\mathrm{loss}(\omega)$ from the magnon-photon coupling includes two parts: $\Sigma^\mathrm{gain/loss}(\omega)=g^2/[i(\omega-\omega_\mathrm{s})\pm\beta]$ for gain ($+$) and loss ($-$), respectively. We note that $\Sigma(\omega)$ now is purely imaginary, leading to a fully transparent transmission at resonance.

\begin{figure}
  \centering
  \includegraphics[width=\linewidth]{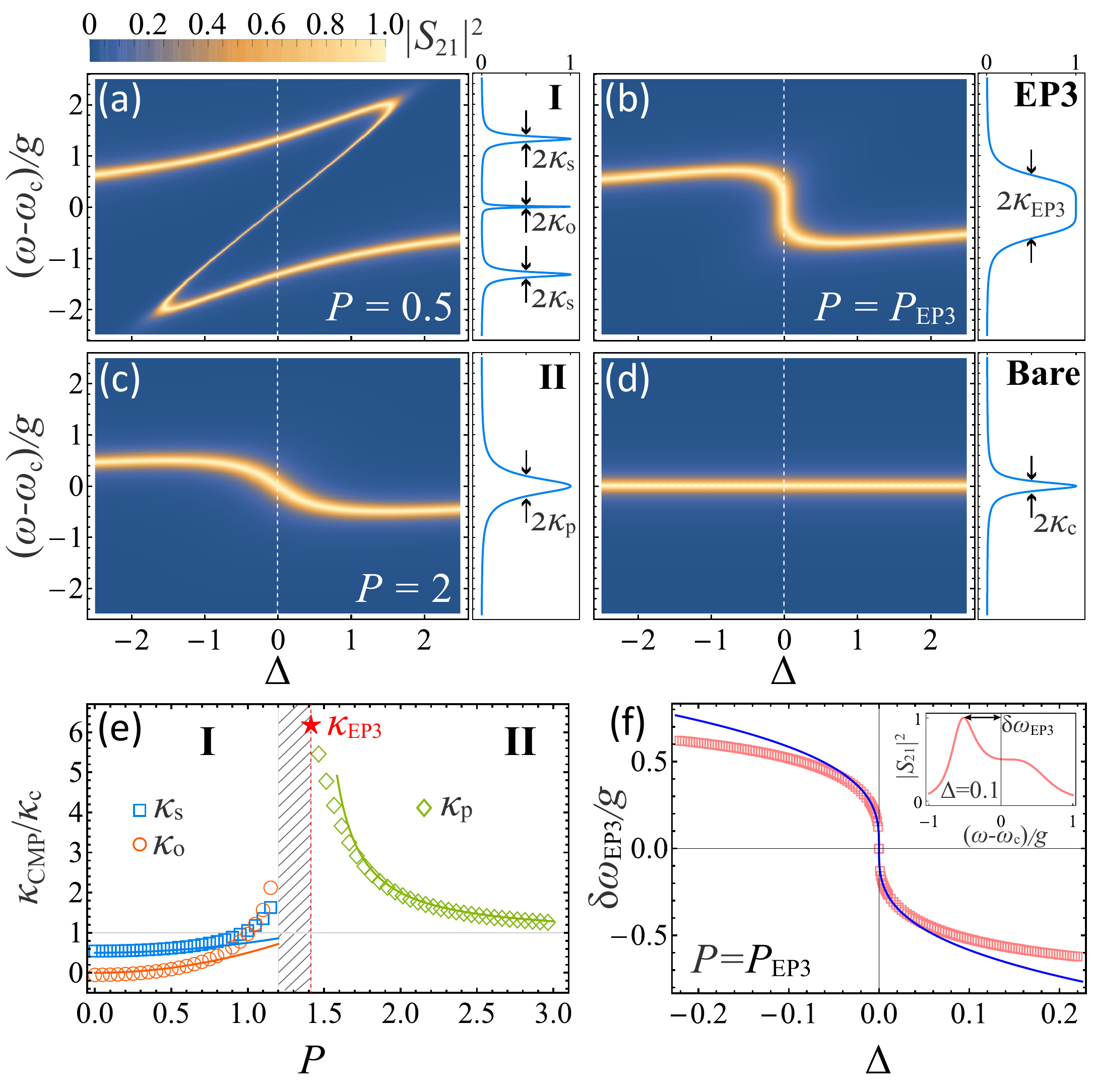}\\
  \caption{
  Transmission spectrum for different gain-loss parameters: (a) $P=0.5$, (b) $P=\sqrt{2}$, and (c) $P=2$. (d) Transmission spectrum of a bare cavity. The right panel in (a)-(d) shows the zero-detuning spectrum. (e) Half-linewidth of CMP modes $\kappa_\mathrm{CMP}\in\{\kappa_\mathrm{s},\kappa_\mathrm{o},\kappa_\mathrm{EP3},\kappa_\mathrm{p}\}$ as a function of the gain-loss parameter $P$ at the zero detuning point. Symbols are numerical results and solid curves are asymptotic formulas (\ref{Eq-width}). Dashed area is not accessible because of the strong overlap between modes. (f) Sensitivity at $P=P_\mathrm{EP3}$. Symbols denote numerical results and the blue curve represents the analytical formula (\ref{Sensitivity}). (Inset) Transmission spectrum as a function of the mode frequency at detuning $\Delta=0.1$. We set $\kappa_{\mathrm{c}}/g=0.1$ in the calculations.}\label{S21}
\end{figure}

Figures \ref{S21}(a)-(c) show the transmission spectrum $|S_{21}|^{2}$ as a function of the mode frequency $\omega$ and the cavity detuning $\Delta$, under different gain-loss parameters $P$. As a reference, we plot the bare-cavity spectrum in Fig. {\ref{S21}}(d). For $P<P_\mathrm{EP3}$, we find that the transmission spectrum displays a novel $``Z"-$shape [see Fig. {\ref{S21}} (a)], instead of the conventional level anti-crossing. Further, we observe three peaks in the strong-coupling region ($\Delta\sim0$), in which the ultra narrow central mode corresponds to the dark-state CMP, beside two sideband abnormal Rabi-splitting modes. To clarify it, we introduce a bright operator $\hat{a}_\mathrm{B}=\frac{1}{\sqrt{2}}(\hat{s}_1+\hat{s}_2)$ and a dark operator $\hat{a}_\mathrm{D}=\frac{1}{\sqrt{2}}(\hat{s}_1-\hat{s}_2)$. The Hamiltonian (\ref{Eq-Ham}) then transforms into
$\mathcal{H}=\hbar\omega_\mathrm{c}\hat{a}^\dagger \hat{a}
  +\hbar\omega_\mathrm{s}\left( \hat{a}_\mathrm{B}^\dagger \hat{a}_\mathrm{B}+\hat{a}_\mathrm{D}^\dagger \hat{a}_\mathrm{D}\right)
  +\hbar \sqrt{2}g\left({a^\dagger}\hat{a}_\mathrm{B}+h.c.\right)+i\hbar\beta\left( \hat{a}_\mathrm{B}^\dagger \hat{a}_\mathrm{D}+h.c.\right)$.
It is straightforward to see that the bright magnon directly couples with the cavity photon, while the dark magnon interacts with the bright magnon through the gain-loss term. States $\left\{|C\rangle\equiv \hat{a}^\dagger|0\rangle, ~ |B\rangle\equiv \hat{a}^\dagger_\mathrm{B}|0\rangle,~|D\rangle\equiv \hat{a}^\dagger_\mathrm{D}|0\rangle\right\}$ now form the new complete basis. At zero detuning, the eigenstate of the central mode is $\frac{i\sqrt{2}}{P}|D\rangle+|C\rangle$, which is totally decoupled from the bright mode $|B\rangle$. We therefore call it dark-state CMP, which may have applications on frequency stabilizations and high-resolution spectroscopic measurements \cite{Lukin1998}. Unlike the conventional dark state with an infinitely long lifetime, the lifetime of the dark-state CMP here is determined by the gain-loss mechanism. Away from the zero-detuning point, its linewidth increases biquadraticlly with $|\Delta|$ (not shown).

An increasing $P$ leads to a coalescence of the peaks. For $P=P_\mathrm{EP3}$, three eigenvalues merge together at $\omega_\mathrm{c}$, and form a flat and wide transparent window shown in Fig. {\ref{S21}}(b). When $P$ further increases, i.e., $P>P_\mathrm{EP3}$, the dark-state CMP mode still survives, with its linewidth however being significantly broadened as plotted in Fig. {\ref{S21}}(c). It is a Purcell-like effect induced by the $\mathcal{P} \mathcal{T}$ symmetry breaking. We are interested in the $P$-dependence of the spectrum linewidth under a zero detuning, and derive the following asymptotic formulas [solid curves in Fig. {\ref{S21}}(e)],
\begin{subequations}\label{Eq-width}
\begin{eqnarray}
P\to0&:&~\kappa_\mathrm{s}/\kappa_\mathrm{c}\simeq (P^2+2)/4,~~~~ \kappa_\mathrm{o}/\kappa_\mathrm{c}\simeq P^2/2,\\
P\gg P_\mathrm{EP3}&:&~\kappa_\mathrm{p}/\kappa_\mathrm{c}\simeq P^2/(P^2-2),
\end{eqnarray}
\end{subequations}
which agree well with numerical results [symbols in Fig. {\ref{S21}}(e)].

\begin{center}
\textbf{C. Magnetic sensitivity }
\end{center}

It has been shown that the non-Hermitian degeneracy can provide an enhancement of sensitivity $\propto|\Delta|^{1/N}$ \cite{Wiersig2014,Hodaei2017,Chen2017} at the $N-$th order EP. The sensitivity is conventionally defined as the splitting of eigenfrequencies perturbed around the EP. However, it becomes unfeasible due to the significant spectrum broadening near the EP as shown in Fig. {\ref{S21}}(b). Further, due to the complex nature of the frequency bifurcation in the vicinity of EP, the view of exceptional precision of exceptional-point sensors has been challenged \cite{Langbein2018,Mortensen2018} by arguing that the sensitivity of EP2 is limited by quantum fluctuations \cite{Langbein2018} and/or statistical noises \cite{Mortensen2018}. In the present model, there always exists a real central mode no matter whether the $\mathcal {P}\mathcal {T}$ symmetry is broken or not. We therefore suggest a more appropriate definition of the sensitivity as the separation between the always-real central mode and the constant cavity mode. At $P=P_\mathrm{EP3}$,
\begin{equation}\label{Sensitivity}
    \delta\omega_\mathrm{EP3}/g=-\mathrm{sgn}(\Delta)\delta \theta, ~~~~
    \text{with}\ \ \delta \theta=2^{1/3}|\Delta|^{1/3},
\end{equation}
excellently consistent with numerical results in the small detuning regime, as plotted in Fig. {\ref{S21}}(f). It can be found that within a detuning $|\Delta|<0.06$ (corresponding to the magnetic field $|\delta B|<2$ mT for $g\sim1$ GHz), our prediction of the exceptional sensitivity of the order $|\Delta|^{1/3}$ is still pronounced. It is noted that our theory is based on the exact $\mathcal {P}\mathcal {T}$-symmetric condition, i.e., the gain and loss is balanced. This condition is necessary for obtaining the third-order exceptional point and for the ultrahigh sensitivity. How about the exact equality does not hold? In such case, the CMP system cannot reach a steady state. A practical experimental measurement therefore should be implemented within the lifetime of the CMP which can be estimated to be $\tau\sim1/|\beta-\beta'|$ with $\beta$ and $\beta'$ the gain and loss coefficient, respectively. Within this timescale $\tau$, our result is still valid.

Sample fluctuations are ubiquitous. To investigate the fluctuation effect on the sensing performance, we consider an ensemble of CMP systems at EP3 with a Gaussian distribution of the detuning parameter $\Delta$,
\begin{equation}
    W(\Delta-\Delta_0)=\frac{1}{\sqrt{2\pi}\sigma}
    \exp\left[-\frac{1}{2}(\Delta-\Delta_0)^2/\sigma^2\right],
\end{equation}
with the target detecting signal at $\Delta_0$ and the noise $\sigma$. For simplicity, we assume $\Delta_0\geqslant0$. The ensemble-average sensitivity is then
\begin{equation}
    \langle \delta\theta \rangle=\frac{\sigma^{1/3}}{2^{1/6}\sqrt{\pi}}\int^{\infty}_{-\infty} |x+x_0|^{1/3}e^{-\frac{1}{2}x^2}\mathrm{d}x,
\end{equation}
\begin{figure}[!htbp]
  \centering
  \includegraphics[width=0.8\linewidth]{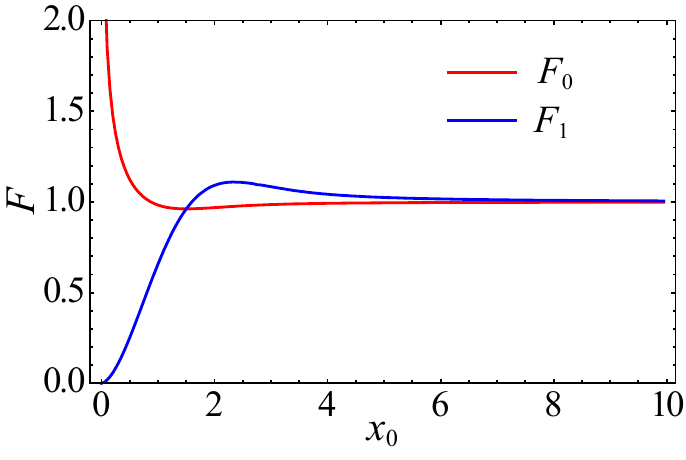}\\
  \caption{Sensitivity-diminution factor as a function of $x_0$.}\label{fluctuations}
\end{figure}
with $x_0=\Delta_0/\sigma$, which in the small and large signal/noise ratio limits reduces to
\begin{equation}
  \langle \delta\theta \rangle\simeq
  \left\{\begin{array}{cc}
                  \sqrt{\frac{2}{\pi}}~\Gamma\left(\frac{2}{3}\right)\sigma^{1/3}, &~~~ x_0\ll 1, \\[1em]
                  2^{1/3}\Delta_0^{1/3}, &~~~ x_0\gg1.
                \end{array}
  \right.
\end{equation}

The sensitivity is found to be free from noise under a large signal/noise ratio, i.e., $x_0\gg1$, while it is suppressed by fluctuations in the opposite limit. To see how fast the sensitivity is recovered, we introduce a sensitivity-diminution factor $F_0= 2^{-1/3}\Delta_0^{-1/3}\langle \delta\theta \rangle$, so that $F_0=1$ represents the noise-free sensitivity regime. The red curve in Fig. \ref{fluctuations} clearly shows that the noise-less sensing is well performed for $x_0\gtrsim1$. In Ref. \cite{Mortensen2018}, a different definition of the sensitivity however is introduced as $\partial \langle \delta\theta \rangle/\partial \Delta_0$ with the following asymptotic form
\begin{equation}
  \frac{\partial \langle \delta\theta \rangle}{\partial \Delta_0}\simeq
  \left\{\begin{array}{cc}
                  \frac{\Gamma\left(\frac{5}{3}\right)}{\sqrt{2\pi}\sigma^{5/3}}\Delta_0, &~~~ x_0\ll 1, \\[1em]
                  \frac{2^{1/3}}{3}\Delta_0^{-2/3}, &~~~ x_0\gg1.
                \end{array}
  \right.
\end{equation}
Similarly, we introduce $F_1=\frac{3}{2^{1/3}}\Delta_0^{2/3}\frac{\partial \langle \delta\theta \rangle}{\partial \Delta_0}$ being the sensitivity-diminution function (see the blue curve in Fig. \ref{fluctuations}). There is a clear diminution of the sensitivity to the noise fluctuations for $x_0\ll1$.
For both cases, we find the condition for the noise-less sensing performance as $x_0\gtrsim2$. Such features near the $3$rd order EP can be utilized for designing magnetic sensor with very high precisions. Considering the cavity frequency resolution $|\delta \omega_\mathrm{EP3}| \sim \kappa_\mathrm{c}$, we obtain the magnetic sensitivity
\begin{equation}
    |\delta B|\approx \frac{\kappa_\mathrm{c}}{2\gamma C},
\end{equation}
where $\gamma$ is the gyromagnetic ratio and $C\sim g^2/\kappa_\mathrm{c}^2$ is the strong coupling cooperativity ranging from $10^3\sim 10^7$ \cite{Huebl2013,Zhang2014,Goryachev2014}. For a microwave cavity working at GHz with a MHz resolution and a (sub-)MHz noise, we estimate the sensitivity $\sim10^{-15}\ \mathrm{T}\ \mathrm{Hz}^{-1/2}$, which is two orders of magnitude
higher than that of the state-of-the-art magnetoelectric sensors \cite{Annapureddy2017}.

\begin{center}
\textbf{D. Nonlinearity }
\end{center}

Nonlinear effects have been completely ignored in the above calculation, which is justified only when the average magnon number is negligibly small. At a mean-field level, the nonlinear correction can be taken into account by modifying the magnon part of Eqs. (\ref{Eq-Lang}) as,
\begin{subequations}
\begin{eqnarray}
&&\dot{\hat{s}}_1 =\left(-i\omega_\mathrm{s}-i\eta\langle \hat{s}_1^\dagger\hat{s}_1\rangle+\beta\right)\hat{s}_1-ig\hat{a}, \hspace{2em}\mathrm{(gain)}\\
&&\dot{\hat{s}}_2 =\left(-i\omega_\mathrm{s}-i\eta\langle \hat{s}_2^\dagger\hat{s}_2\rangle-\beta\right)\hat{s}_2-ig\hat{a}, \hspace{2em}\mathrm{(loss)}
\end{eqnarray}
\end{subequations}
where $\eta$ is the nonlinear coefficient from the magnetic anisotropy \cite{Yan2017,Wang2018}. Solving these equations, we obtain $\langle \hat{s}_1^\dagger\hat{s}_1\rangle=\langle \hat{s}_2^\dagger\hat{s}_2\rangle=\frac{n_\mathrm{p}}{P^2+\Delta^2}$, with $n_\mathrm{p}=\langle\hat{a}^\dagger\hat{a}\rangle$ the
average photon number in the cavity. For a mall $\eta$, the EP3 is slightly shifted to $\Delta=-\frac{\eta n_\mathrm{p}}{2}$ with $P=\sqrt{2}$.

\begin{center}
\textbf{IV. WAVE SCATTERING CALCULATION }
\end{center}

\begin{figure}
  \centering
  \includegraphics[width=\linewidth]{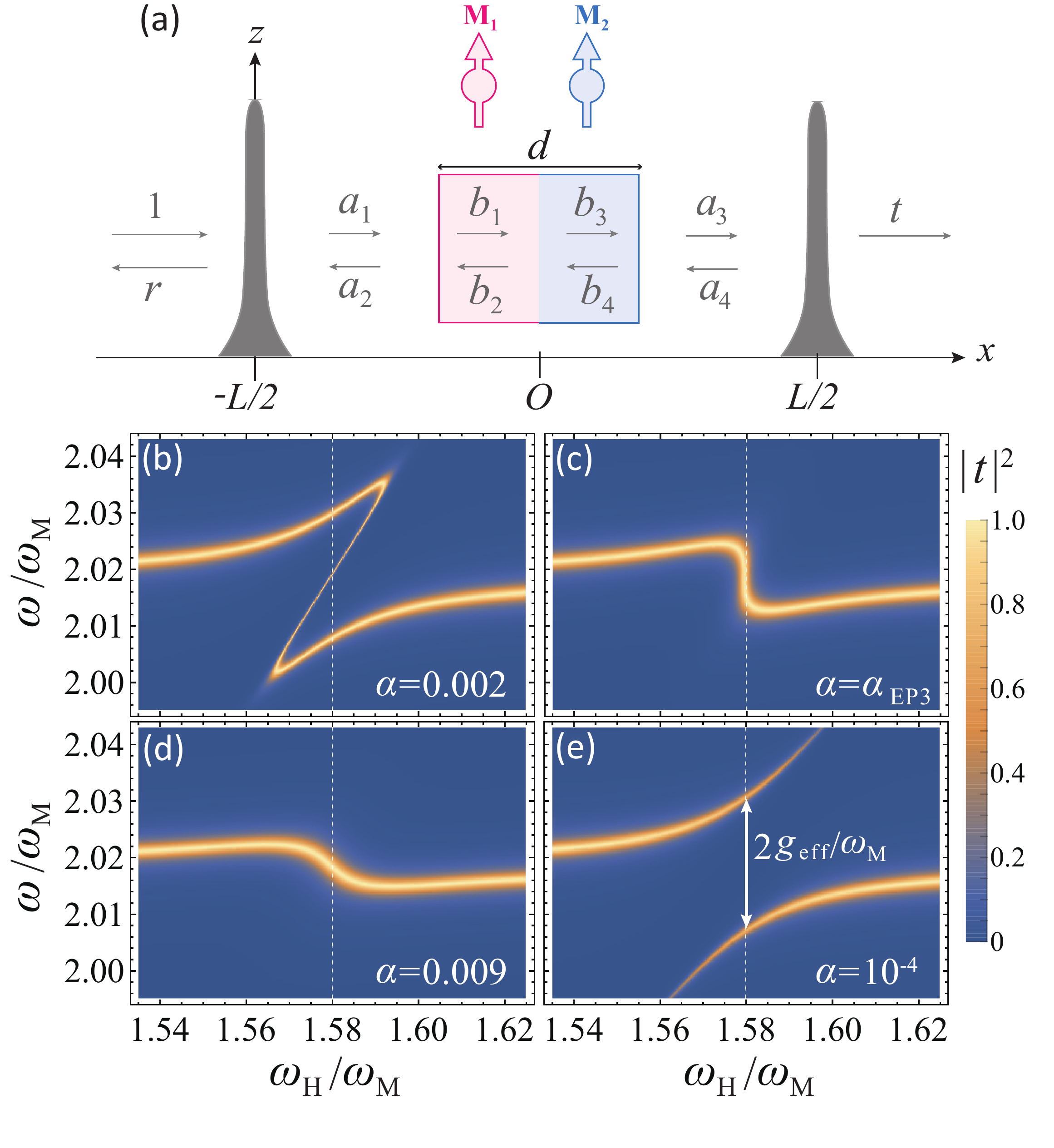}\\
  \caption{(a) Schematic plot of 1-dimensional scattering model for a $\mathcal{P} \mathcal{T}$-symmetric magnetic bilayer in a microwave cavity.
  Transmission spectrum (b) in the $\mathcal{P}\mathcal{T}$ symmetric region,
  (c) at the exceptional point, and (d) in the $\mathcal{P}\mathcal{T}$ broken phase.
  (e) Conventional anti-crossing spectrum with $\alpha_1=\alpha_2=\alpha=10^{-4}$.
  The following parameters are adopted: $\ell=L=46$ mm and $d=5~\mu$m. The cavity resonant frequency of interest is then $\omega_\mathrm{c}/\omega_\mathrm{M}\simeq2.019$. The external magnetic field is $\omega_\mathrm{H}/\omega_\mathrm{M}\simeq1.58$ at the split point.}\label{scattering}
\end{figure}

So far we illustrated the essence of $\mathcal{P} \mathcal{T}$-symmetric CMPs only through a toy model Hamiltonian (\ref{Eq-Ham}). The single-particle assumption adopted in the quantum Hamiltonian formalism and the conclusions accordingly needs justifications. Below, we explicitly show that the major results are valid in the classical limit which in principle includes multi-particle effects. Further, a physical realization is necessary to be sought to testify the theoretical predictions. To this end, we follow the one-dimensional scattering method in Ref. \cite{Cao2015}, and consider a ferromagnetic bilayer placed in a microwave cavity [as shown in Fig. \ref{scattering}(a)]. The cavity wall is modelled by a delta permeability function $\mu=\mu_0\left[ 1+2\ell\delta(x+L/2)+2\ell\delta(x-L/2)\right]$, where $L$ is the cavity width and $\ell$ is the wall opacity.
The dynamics of magnetization $\mathbf{M}$ is governed by the Landau-Lifshitz-Gilbert (LLG) equation,
\begin{equation}
\frac{\partial\mathbf{M}_{j}}{\partial t}=-\gamma \mu_0 \mathbf{M}_{j}\times\mathbf{H}_{\mathrm{eff}}%
+\frac{\alpha_{j}}{M_{\mathrm{s}}}\mathbf{M}_{j}\times\frac{\partial\mathbf{M}_{j}}{\partial t},
\label{Eq-LLG}%
\end{equation}
where $\mu_0$ is the vaccum permeability. The effective magnetic field $\mathbf{H}_{\mathrm{eff}}%
=H\hat{z}+\mathbf{h}$ consists of the external and rf magnetic fields. $\mathbf{M}_{j}$ with $j=1,2$ labels the left and right magnets with balanced magnetic gain and loss $\alpha_1=-\alpha$ and
$\alpha_2=\alpha~(\alpha>0)$, respectively. The typical value of $\alpha$ ranges from $10^{-5}$ to $10^{-1}$.

For small-amplitude magnetization oscillations $\mathbf{M}_{j}=M_{\mathrm{s}}\hat{z}+\mathbf{m}_{j}$ with $|\mathbf{m}_{j}|\ll M_{\mathrm{s}}$ and $M_{\mathrm{s}}$ being the saturation magnetization, $\mathbf{m}_{j}$ is driven by the rf magnetic field $\mathbf{h}$ satisfying Maxwell's equation
\begin{equation}
\left(  \nabla^{2}+k_{\varepsilon}^{2}\right)  \mathbf{h}=\nabla
(\nabla\cdot\mathbf{h})-k_{\varepsilon}^{2}\mathbf{m}, \label{Eq-Maxs}%
\end{equation}
where $k_{\varepsilon}^{2}=\varepsilon\mu_{0}\omega^{2}=\varepsilon_\mathrm{r} q^{2}$, $q$ is the vacuum light wave-vector, $\varepsilon_\mathrm{r}=\varepsilon/\varepsilon_{0}$ is the relative permittivity of ferromagnets, and $\mathbf{m}=\mathbf{m}_{1(2)}$ for $-d/2\leqslant x<0 \ (0<x\leqslant d/2)$.

Assuming a linearly polarized microwave field $h_y(x,t)=\psi(x)e^{-i\omega t} $ traveling along the $\hat{x}-$direction,
the wave vector in magnetic bilayer takes the form $k_j=q\sqrt{\varepsilon_\mathrm{r}\mu_{\mathrm{v},j}}$ \cite{Cao2015} for a given frequency $\omega$,
where $\mu_{\mathrm{v},j}=\frac{\omega^2-(\omega_j+\omega_\mathrm{M})^2}{\omega^2-\omega_j(\omega_j+\omega_M)}$
is the Voigt permeability with $\omega_j=\omega_\mathrm{H}-i\omega\alpha_j$, $\omega_{\mathrm{H}}=\gamma \mu_0 H$, and $\omega_{\mathrm{M}}=\gamma \mu_0 M_{\mathrm{s}}$.
The microwave field $\psi(x)$ in different regimes [see Fig. \ref{scattering}(a)] can be expressed as
\begin{subequations}
\begin{eqnarray}
    &&\psi_1 = e^{iqx}+r e^{-iqx},~~~ \psi_2 = a_1 e^{iqx}+a_2 e^{-iqx},\\
    &&\psi_3 = b_1 e^{ik_1x}+b_2 e^{-ik_1x},~~~ \psi_4 = b_3 e^{ik_2x}+b_4 e^{-ik_2x},\\
    &&\psi_5 = a_3 e^{iqx}+a_4 e^{-iqx},~~~ \psi_6 = t e^{iqx},
\end{eqnarray}
\end{subequations}
where coefficients $\{r,t,a_{1},a_{2}, a_3, a_4, b_{1},b_{2},b_3,b_4\}$ are determined
by the electromagnetic boundary conditions at the interfaces.

We adopt the magnetic material parameters of YIG in the calculations, e.g., $\varepsilon_\mathrm{r}=15$ \cite{Sadhana2009}, $\mu_{0}M_{\mathrm{s}}=175\,\mathrm{mT}$ \cite{Manuilov2009}, and $\gamma/(2\pi)=28\,\mathrm{GHz/T}$. Figure \ref{scattering}(b) shows the transmission spectrum for a small damping $\alpha=0.002$, which exhibits a similar $``Z"-$shape with Fig. \ref{S21}(a). From the wave-scattering calculation, we establish the following correspondence of parameters in the toy model and the present one
\begin{equation}
   \omega_\mathrm{s}\!=\!\sqrt{\omega_\mathrm{H}(\omega_\mathrm{M}+\omega_\mathrm{H})},~~~
   \beta=\frac{\alpha}{2}\sqrt{\omega_\mathrm{M}^2+4\omega_\mathrm{c}^2},~~~g=\frac{g_\mathrm{eff}}{\sqrt{2}},
\end{equation}
where $g_\mathrm{eff}$ is the effective coupling strength represented by the anti-crossing gap of the conventional strong-coupling spectrum \cite{Cao2015} [see also Fig. \ref{scattering}(e)]. For a ferromagnetic bilayer of the thickness $d=5~\mu$m, its value can be found from the spectrum $g_\mathrm{eff}\simeq0.012\omega_\mathrm{M}$. We therefore deduce the critical Gilbert-type gain-loss parameter $\alpha_\mathrm{EP3}\simeq0.0057$. The transmission spectrum at the EP3 is plotted in Fig. \ref{scattering}(c), which demonstrates similar dependence on the frequency and the detuning as Fig. \ref{S21}(b). For a large $\alpha$, the system goes into the $\mathcal{P} \mathcal{T}$ symmetry broken phase, with a Purcell-like effect shown in Fig. \ref{scattering}(d).

\begin{center}
\textbf{V. DISCUSSION}
\end{center}

From an experimental point of view, negative magnetic damping is necessary for observing the $\mathcal{P} \mathcal{T}-$symmetric CMP. Although a natural gain in magnetic materials may not be realistic, an effective negative damping can be realized by, for instance, parametric driving from an ac magnetic field \cite{Lee2015}. Spin transfer torques can be either parallel or antiparallel to the intrinsic damping depending on the the current direction. A sufficiently large current density over a critical value can result in an amplification of magnetization oscillation, thus realizing a magnetic gain \cite{Braganca2005,Ralph2008,Lee2015}. Wegrowe \emph{et al} systematically investigated the spin transfer in an open ferromagnetic system, and found that the negative damping emerges naturally for describing the spin exchange between the magnet and the environment \cite{Wegrowe2007}. A recent experiment demonstrated that the electric-field can induce a negative magnetic damping in heterostructured ferroelectric$|$ferromagnet layers \cite{Jia2015}. Parametric coupling can induce a Gilbert-like gain by an optical laser \cite{Kusminskiy2016}. To soften the experimental challenge for realizing a negative magnetic damping, we propose a $\mathcal{P} \mathcal{T}-$symmetric synthetic electric circuit coupled with a ferromagnetic sphere to implement our model (see Fig. \ref{circuit} and details in Appendix B).

\begin{center}
\textbf{VI. CONCLUSIONS }
\end{center}

To conclude, we predicted an ultrahigh magnetic sensitivity of $\mathcal{P} \mathcal{T}-$symmetric cavity magnon polaritons near the third-order EP. The estimated sensitivity approaches fetotesla, and is not limited by the quantum or statistical noise under proper conditions. Higher cooperativity of coupled magnon and photon hybrid can further improve the sensing performance. We propose to use magnetic bilayers with balanced gain and loss in a microwave cavity or $\mathcal{P} \mathcal{T}-$symmetric circuits coupled with a magnetic sphere to experimentally verify our predictions. This study provides the theoretical framework for the emerging $\mathcal {P}\mathcal {T}-$symmetric spin cavitronics, and offers a new pathway for designing ultrasensitive magnetometers. A generalization of present results to arbitrarily high-order exceptional points should be an interesting issue for future study.

\begin{center}
\textbf{ACKNOWLEDGMENTS }
\end{center}

This work was supported by the National Natural Science Foundation of China (Grants No. 11704060 and No. 11604041), the National Key Research Development Program under Contract No. 2016YFA0300801, and the National Thousand-Young-Talent Program of China.

\begin{center}
\textbf{APPENDIX }
\end{center}
\begin{center}
\textbf{A. Input-output formalism }
\end{center}
\setcounter{equation}{0}
\renewcommand{\theequation}{A.\arabic{equation}}

An input-output theory \cite{Gardiner1985} is derived for a cavity interacting with a thermal bath. Our starting point is the total Hamiltonian
\begin{equation}
 \mathcal{H}_\mathrm{total}=\mathcal{H}_\mathrm{sys}+\mathcal{H}_\mathrm{bath}+\mathcal{H}_\mathrm{int},
\end{equation}
where $\mathcal{H}_\mathrm{sys}$ describes the intracavity dynamics, the same as Eq. (1) in the main text, $\mathcal{H}_\mathrm{bath}$ is the bath Hamiltonian
\begin{equation}
  \mathcal{H}_\mathrm{bath}=\hbar\sum_k \omega_k \hat{b}_k^\dagger\hat{b}_k,
\end{equation}
with the bosonic creation (annihilation) operator $\hat{b}_k^\dagger$ ($\hat{b}_k$).
They are coupled by the interaction term
\begin{equation}
  \mathcal{H}_\mathrm{int}=\hbar \sum_k  \left(f_k\hat{a}^\dagger\hat{b}_k+ f_k^* \hat{b}_k^\dagger\hat{a}\right),
\end{equation}
with the commutation relation $[\hat{a}, \hat{a}^\dagger]=1$ and $[\hat{b}_k, \hat{b}_{k'}^\dagger]=\delta_{kk'}$. $f_k$ is the coupling strength.

In Heisenberg picture,
the time-dependent operator $\hat{\mathcal{O}}(t):=e^{i\mathcal{H}t}\hat{\mathcal{O}}e^{-i\mathcal{H}t}$ satisfies the
following equations
\begin{subequations}
\begin{eqnarray}\label{Eq-a}
    &&\dot{\hat{a}}(t) =\frac{i}{\hbar}\left[\mathcal{H}_\mathrm{sys},\hat{a}(t)\right]-i\sum_kf_k \hat{b}_k(t),\\
    &&\dot{\hat{b}}_k(t)=-i\omega_k\hat{b}_k(t)-if_k^*\hat{a}(t).
\end{eqnarray}
\end{subequations}
The formal solution of $\hat{b}_k(t)$ can be written as
\begin{subequations}
\begin{eqnarray}
    \hat{b}_k(t)=\hat{b}_k(t_0)e^{-i\omega_k(t-t_0)}-if_k^*\int_{t_0(<t)}^t \!\mathrm{d} \tau e^{-i\omega_k(t-\tau)} \hat{a}(\tau),\\
    \hat{b}_k(t)=\hat{b}_k(t_1)e^{-i\omega_k(t-t_1)}+if_k^*\int_{t}^{t_1(>t)}\!\mathrm{d} \tau e^{-i\omega_k(t-\tau)} \hat{a}(\tau).
\end{eqnarray}
\end{subequations}
We thus have
\begin{eqnarray}
    \dot{\hat{a}}(t) &=&\frac{i}{\hbar}\left[\mathcal{H}_\mathrm{sys},\hat{a}(t)\right]
        -i\sum_kf_k \hat{b}_k(t_0)e^{-i\omega_k(t-t_0)} \nonumber \\
    &&-\sum_k|f_k|^2\int_{t_0}^t\mathrm{d} \tau e^{-i\omega_k(t-\tau)} \hat{a}(\tau),~~~~(t_0<t)\\
    &=&\frac{i}{\hbar}\left[\mathcal{H}_\mathrm{sys},\hat{a}(t)\right]
       -i\sum_kf_k \hat{b}_k(t_1)e^{-i\omega_k(t-t_1)}\nonumber \\
    &&+\sum_k|f_k|^2\int_{t}^{t_1}\mathrm{d} \tau e^{-i\omega_k(t-\tau)} \hat{a}(\tau).~~~~(t<t_1)
\end{eqnarray}

We then aim to convert the summation to the integral by introducing the mode density $\rho_k$. Assuming that both the mode density $\rho_k$ and
the coupling strength $f_k$ are mode independent, i.e., $\rho_k=\rho$ and $f_k=f$, we obtain the following relation
\begin{eqnarray}
    \sum_k\mapsto \int\rho\mathrm{d}\omega_k, &~&
    \kappa_\mathrm{c} = 2\pi\rho|f|^2,\nonumber\\
    \int_{-\infty}^{\infty}\mathrm{d}\omega_k e^{-i\omega_k(t-t')}&=&2\pi\delta(t-t'),\nonumber\\
\int_{t_0}^{t}\!\mathrm{d}\tau\delta(t-\tau)\hat{a}(\tau)=\int_{t}^{t_1}\!\!\!\!&\mathrm{d}\tau&\!\!\! \delta(t-\tau)\hat{a}(\tau)=\frac{1}{2}\hat{a}(t).
\end{eqnarray}
The input and output fields are defined as
\begin{subequations}
\begin{eqnarray}
    \hat{b}_\mathrm{in}(t)\equiv \frac{i}{\sqrt{\kappa_\mathrm{c}}}\sum_k f_k\hat{b}_k(t_0)e^{-i\omega_k(t-t_0)},\\
    \hat{b}_\mathrm{out}(t)\equiv\frac{i}{\sqrt{\kappa_\mathrm{c}}}\sum_k f_k\hat{b}_k(t_1)e^{-i\omega_k(t-t_1)}.
\end{eqnarray}
\end{subequations}
The coefficient $1/\sqrt{\kappa_\mathrm{c}}$ in the expressions guarantees that the input and output fields satisfy the bosonic commutation relations
\begin{equation}
    \left[\hat{b}_\mathrm{in}(t), \hat{b}^\dagger_\mathrm{in}(t')\right]=
    \left[\hat{b}_\mathrm{out}(t), \hat{b}^\dagger_\mathrm{out}(t')\right]=\delta(t-t').
\end{equation}
Then, Eq. (\ref{Eq-a}) can be simplified to
\begin{eqnarray}
    \dot{\hat{a}}(t) &=&\frac{i}{\hbar}\left[\mathcal{H}_\mathrm{sys},\hat{a}(t)\right]-\sqrt{\kappa_c}\hat{b}_\mathrm{in}(t)-\frac{\kappa_\mathrm{c}}{2}\hat{a}(t)~~~~\\
     &=&\frac{i}{\hbar}\left[\mathcal{H}_\mathrm{sys},\hat{a}(t)\right]-\sqrt{\kappa_c}\hat{b}_\mathrm{out}(t)+\frac{\kappa_\mathrm{c}}{2}\hat{a}(t),
\end{eqnarray}
from which we obtain the input-output formula \cite{Meystre2007,Walls2008},
\begin{equation}
   \hat{b}_\mathrm{out}(t)= \hat{b}_\mathrm{in}(t)+\sqrt{\kappa_\mathrm{c}}\hat{a}(t).
\end{equation}

For a two-port cavity, the input and output fields are connected by a scattering matrix,
\begin{equation}
    \left(
      \begin{array}{c}
        \hat{b}_\mathrm{out}^{(1)} \\[0.5em]
        \hat{b}_\mathrm{out}^{(2)} \\
      \end{array}
    \right)
    =\left(
       \begin{array}{cc}
         S_{11} & S_{12} \\[0.5em]
         S_{21} & S_{22} \\
       \end{array}
     \right)
     \left(
       \begin{array}{c}
         \hat{b}_\mathrm{in}^{(1)} \\[0.5em]
         \hat{b}_\mathrm{in}^{(2)} \\
       \end{array}
     \right).
\end{equation}
So, every port satisfies
the input-output relation $\hat{b}_\mathrm{out}^{(1,2)}=\hat{b}_\mathrm{in}^{(1,2)}+\sqrt{\kappa_\mathrm{c}} \hat{a}$, while the total field satisfies
$\hat{b}_\mathrm{out}=\hat{b}_\mathrm{in}+2\sqrt{\kappa_\mathrm{c}} \hat{a}$ with the total input field $\hat{b}_\mathrm{in}=\hat{b}_\mathrm{in}^{(1)}+\hat{b}_\mathrm{in}^{(2)}$,
and total output field $\hat{b}_\mathrm{out}=\hat{b}_\mathrm{out}^{(1)}+\hat{b}_\mathrm{out}^{(2)}$. Considering
only one input field from port 1, i.e., $\hat{b}_\mathrm{in}^{(2)}=0$, we then obtain the quantum Langevin equations,
\begin{eqnarray}
    &&\dot{\hat{a}}=(-i\omega_\mathrm{c}-\kappa_\mathrm{c})\hat{a}-ig(\hat{s}_1+\hat{s}_2)-\sqrt{\kappa_\mathrm{c}} \hat{b}_\mathrm{in},\nonumber\\
    &&\dot{\hat{s}}_1 =(-i\omega_\mathrm{s}+\beta)\hat{s}_1-ig\hat{a}, \nonumber\\
    &&\dot{\hat{s}}_2 =(-i\omega_\mathrm{s}-\beta)\hat{s}_2-ig\hat{a}.
\end{eqnarray}
Solving the above equations in frequency space, we obtain,
\begin{equation}
    a(\omega)=\frac{\sqrt{\kappa_\mathrm{c}} b_\mathrm{in}(\omega)}{i(\omega-\omega_\mathrm{c})-\kappa_\mathrm{c}+\Sigma(\omega)}.
\end{equation}
where $\Sigma(\omega)$ is the self-energy from the magnon-photon coupling including gain and loss parts:
\begin{eqnarray}
\Sigma(\omega)&=&\Sigma^\mathrm{gain}(\omega)+\Sigma^\mathrm{loss}(\omega)\nonumber\\
&=&\frac{g^2}{i(\omega-\omega_\mathrm{s})+\beta}+\frac{g^2}{i(\omega-\omega_\mathrm{s})-\beta}.
\end{eqnarray}
By substituting the above relations into the input-output formula, we have
\begin{eqnarray}
    && b_\mathrm{out}^{(1)}=b_\mathrm{in}^{(1)}
       +\frac{\kappa_\mathrm{c} b_\mathrm{in}^{(1)}}{i(\omega-\omega_\mathrm{c})-\kappa_\mathrm{c}+\Sigma(\omega)},\nonumber\\
    && b_\mathrm{out}^{(2)}=\frac{\kappa_\mathrm{c}b_\mathrm{in}^{(1)}}
      {i(\omega-\omega_\mathrm{c})-\kappa_\mathrm{c}+\Sigma(\omega)}.
\end{eqnarray}
One therefore obtains the frequency-resolved reflection and transmission coefficients,
\begin{eqnarray}
    &&S_{11}=\frac{b_\mathrm{out}^{(1)}}{b_\mathrm{in}^{(1)}}
      =1+\frac{\kappa_\mathrm{c}}{i(\omega-\omega_\mathrm{c})-\kappa_\mathrm{c}+\Sigma(\omega)},\nonumber\\
    &&S_{21}=\frac{b_\mathrm{out}^{(2)}}{b_\mathrm{in}^{(1)}}
     =\frac{\kappa_\mathrm{c}}{i(\omega-\omega_\mathrm{c})-\kappa_\mathrm{c}+\Sigma(\omega)}.
\end{eqnarray}

\begin{center}
\textbf{B. Synthetic electric circuits with $\mathcal{P} \mathcal{T}$ symmetry}
\end{center}
\setcounter{equation}{0}
\renewcommand{\theequation}{B.\arabic{equation}}

We propose a synthetic circuit consisting of two resistance-inductor-capacitor (RLC) resonators with balanced gain and loss coupled to a precessional magnetic sphere, shown in Fig. \ref{circuit}(a). The gain in circuit can be realized through negative resistances \cite{Chua1980,Schindler2012}.
\begin{figure}[!htbp]
  \centering
  \includegraphics[width=0.95\linewidth]{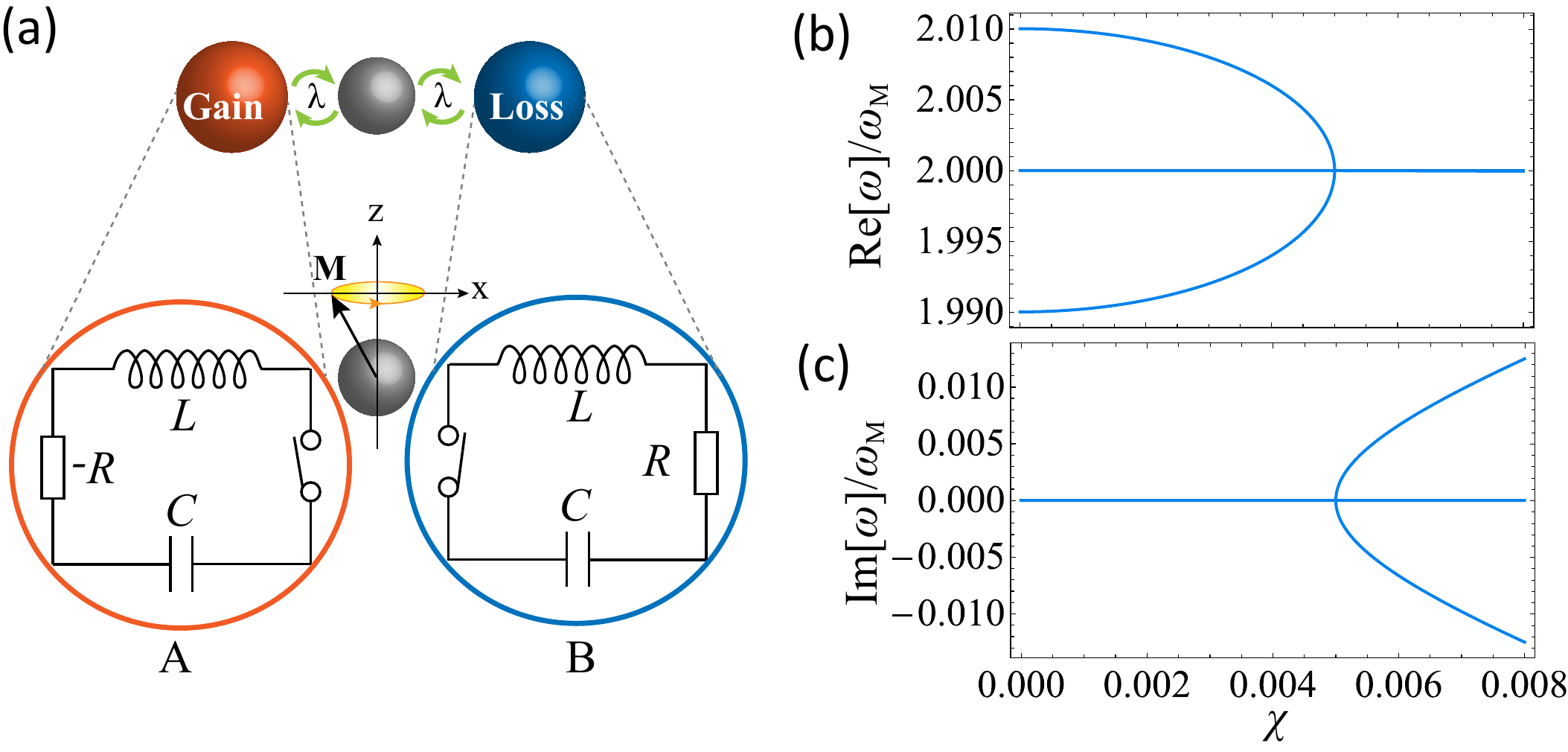}\\
  \caption{(a) Schematic of two resistance-inductor-capacitor (RLC) circuits with balanced gain (red) and loss (blue) inductively coupled to a magnetic sphere with precessing magnetization. (b) Real and (c) imaginary parts of eigenvalues as a function of the gain/loss parameter. EP3 emerges at $\chi_\mathrm{EP3}=0.005$. Parameters are chosen as $\omega_0/\omega_\mathrm{M}=\omega_\mathrm{H}/\omega_\mathrm{M}=2$ and $\lambda=0.01$ in the calculations.}\label{circuit}
\end{figure}

The equation describing the RLC circuits with two coils parallel to the $\hat{x}-$axis is written as \cite{Grigoryan2018}
\begin{subequations}
\begin{eqnarray}
    L \dot{I}_A-RI_A+(1/C)\int I_A \mathrm{d}t=V_A(t),~~~~~\mathrm{(gain)}\\
    L \dot{I}_B+RI_B+(1/C)\int I_B \mathrm{d}t=V_B(t).~~~~~\mathrm{(loss)}
\end{eqnarray}
\end{subequations}
Such a RLC circuit has a characteristic frequency $\omega_0=1/\sqrt{LC}$. The driving voltage induced by the precessing magnetic moment is given by Faraday's law of induction,
\begin{equation}
 V_A(t)=\lambda_{1} L \dot{m}_x, ~~~~~V_B(t)=\lambda_{1} L \dot{m}_x,
\end{equation}
with $\lambda_{1}$ the coupling strength. The magnetization dynamics in the magnetic sphere is governed by the Landau-Lifshitz equation (the Gilbert damping is ignored here),
\begin{equation}
    \mathbf{\dot{M}}=-\gamma\mu_{0}\mathbf{M}\times\mathbf{H},
\end{equation}
where the total magnetic field is $\mathbf{H}=H\hat{z}+h_A\hat{x}+h_B\hat{x}$, and the rf magnetic fields due to the RLC circuit are given by Ampere's law,
\begin{equation}
  h_A=-\lambda_{2} I_A,~~~~h_B=-\lambda_{2} I_B,
\end{equation}
with a coefficient $\lambda_{2}$. We finally obtain the secular equation,
\begin{equation}
 \left(
   \begin{array}{ccc}
     \omega^2\!-\!\omega^2_0\!-\!2i\chi\omega\omega_0 & \omega^2 \lambda^2 & 0 \\
     \omega_\mathrm{M}\omega_\mathrm{H} & \omega^2\!-\!\omega_\mathrm{H}^2 & \omega_\mathrm{M}\omega_\mathrm{H} \\
     0 & \omega^2 \lambda^2 & \omega^2\!-\!\omega^2_0\!+\!2i\chi\omega\omega_0 \\
   \end{array}
 \right)
 \left(
   \begin{array}{c}
     h_A \\
     m_x \\
     h_B \\
   \end{array}
 \right)=0,
\end{equation}
with $\chi=R/(2L\omega_0)$ the dimensionless gain/loss parameter and $\lambda^2=\lambda_{1}\lambda_{2}$. Solving the above equation, we obtain the $\chi-$dependence of the eigenvalues as shown in Fig. \ref{circuit}. At the zero detuning i.e., $\omega_0=\omega_\mathrm{H}$, we observe the third order exceptional point EP3 when $\chi_\mathrm{EP3}=0.005$.

\end{document}